# Interaction Theory of Hazard-Target System

Ji Ge[1, 2, 3]*, Yu-Yuan Zhang[1, 3], Kai-Li Xu [1, 3]*, Ji-Shuo Li[1, 3], Xi-Wen Yao[1, 3]*, Chun-Ying Wu[2], Shuang-Yuan Li[4], Fang Yan[5, 6], Jin-Jia Zhang[7], Qing-Wei Xu[8]

[1]School of Resources and Civil Engineering, Northeastern University, Shenyang 110819, PR China

[2] School of Resources and Environmental Engineering, Jilin Institute of Chemical Technology, Jilin 132022, PR China

[3]Key Laboratory of Ministry of Education on Safe Mining of Deep Metal Mines, Shenyang 110819, PR China

[4]Information Centre, Jilin Institute of Chemical Technology, Jilin 132022, PR China

[5]School of Resources and Safety Engineering, Central South University, Changsha 410083, PR China

[6]Safety & Security Theory Innovation and Promotion Centre, Central South University, Changsha 410083, PR China

[7]Institute of Public Safety Research, Tsinghua University, Beijing 100084, PR China

[8]College of Information and Management Science, Henan Agricultural University, Zhengzhou 450046, PR China

*Corresponding author. Email: geji@jlict.edu.cn (J.G.); xukaili1965@163.com (K.L.X.)

**Abstract—Major accidents (e.g., the Space Shuttle Challenger disaster in the USA, the Bhopal Disaster in India, Fukushima nuclear accident in Japan, Tianjin Port fire and explosion accident in China) have occurred all over the world. Safety scientists are always trying to understand why these accidents happened and how to prevent these accidents. Accident models and theories form the basis for many safety research fields and practices such as investigation of accidents, design of a safer system and decision making on safety related field. There is no universally accepted model with useful elements relating to understanding accident causation, although many accident causation models exist. Based on STAMP and RMF, we proposed a new theory named the Interaction Theory of Hazard-Target System (ITHTS) that incorporate human, organizational and technological characteristics in the same framework. Accident analysis methods provide the necessary information to analysis the accident in a specific setting. In order to solve the issues that current accident analysis methods still face, we proposed a new systemic accident analysis method based on ITHTS and STPA. We choose Tianjin Port fire and explosion accident in China as a case study to demonstrate the viability of the Interaction Theory of Hazard-target System and the applicability of the new accident analysis method. It is concluded that ITHTS can explain the phenomena in safety practice and the new accident analysis method can be application in the explanation and analysis of major accident.**

Major accidents such as the Bhopal Disaster (*1*), the Chernobyl reactor accident (*2*), the Challenger disaster (*3*), the Fukushima nuclear accident (*4*), and the Tianjin Port fire and explosion accident (*5*), have occurred around the world. In order to understand why these major accidents occurred and how to prevent them in future, hundreds of safety models and accident analysis methods have been developed since the accident proneness model was proposed by Greenwood and Woods in 1919 (*6, 7*). As the theoretical foundation of safety science, safety models and theories (or accident models and accident causation models) provide a theoretical framework of accident analysis and prevention. They form the basis for many safety research and practices, such as investigation of accidents, design of a safer system and decision making in safety-related fields. These models and theories can be divided into two specific categories: older models based on causal-sequence thinking and newer models based on systems thinking (*8*). These newer safety models, such as Systems-Theoretic Accident Model and Processes (STAMP) (*9*), Risk Management Framework (RMF) (*10*), Normal Accident Theory (NAT) (*11*), and Functional Resonance Analysis Model (FRAM) (*12*), have advantages over older models (e.g., domino theory, the energy model) in terms of explanatory power, usefulness, and extent of acceptability (*8*). However, the gap between academic developments and the application of these safety models in practice is still substantial (*8, 13*).

It is nearly impossible to underestimate the importance of establishing safety theory for us to understand why an accident must be the way it is (*14*). However, there is no universally accepted model with useful elements related to understanding accident causation, although many safety models exist. Few safety theories or models explicitly provide insight into causal pathways and mechanisms that operate between levels of work system (*15*). Given the widely accepted belief that accidents are a systems problem and require a whole system approach to elucidate the complex etiologies, little evidence exists that this process is occurring in practice among the applications reviewed (*16*). Safety models, which consider the contributory factors (e.g., the influence of social factors such as legislative, regulatory, and cultural factors) across all levels of a sociotechnical system, can provide "a more complete understanding of how and why accidents occur" (*17*). The limitations of accident causation models suggest that an understanding how systems thinking tenets contribute to accident causation is required in future research (*17*). It is surprising that there is no fundamental theory of safety science that is "rich enough to incorporate human, organizational and technological characteristics in the same framework" (*17, 18*). In the following, a new theory named Interaction Theory of Hazard-Target System (ITHTS) based on STAMP (see "Systems-Theoretic Accident Model and Processes" in Methods) is therefore presented and illustrated by application to the analysis of Tianjin Port fire and exploration accident.

**Results**

**Interaction Theory of Hazard-Target System**

*Sociotechnical system*

Systems are "as pervasive as the universe in which we live" (*19*). Every element of the universe can be viewed as a system (*20*). Several efforts have emerged to capture the nature of what a system represents. Although the definition of "system" is controversial, there is a consensus that a system is more than the sum of its parts. A sociotechnical system is consisted of "a cluster of elements including technology, regulation, cultural meanings, markets, infrastructure, maintenance networks, and supply networks" (*21, 22*). Complex systems are often decomposed into manageable components and then analyzed how they might perform together (*23*). Sociotechnical system can therefore be viewed as a system of systems composed of a social system and a technical system, which interact with each other and fit together (*24, 25,26*). The social system is composed of a human system, organizational system, and sociological system. In general, the social system has three aspects: individual factors (or human factors such as humans or people), organizational factors (e.g., government, company, public, industrial association, agent, labor union) and sociological factors (e.g., law, politics, policy, economics and standards). Technical systems refer to physical systems (including machines, devices, and automation equipment), cyber-physical systems and technological factors (such as information technology, the Internet of Things (IoT), 3D printing technology, and artificial intelligence). For example, a smart manufacturing system can be viewed as a sociotechnical system composed of a human subsystem, technical subsystem and organization subsystem (*27*). In fact, the role of humans in the supervisory control of complex, near-automated systems (such as multi-agent systems and cyber-physical production systems) has long been a critical issue (*27*). Every hierarchical level must impose safety constraints on the activity of the level beneath it to control the lower-level systems' behavior (*9*), that is, interactions between IoT devices, software, and people are interpreted as a kind of social relationship (*28*).

## Safety constraints

In systems theory, constraints are often viewed as "limitations on the behavioral degree of freedom of the system components" (*29*). Constraints represent the acceptable way in which the system can achieve its goals (*29*). Safety constraints limit the behavior of the system to ensure that it operates within safe boundaries (*9, 29*). Firesmith (*30*) considered a "safety constraint" as any constraint that specifies a specific safeguard (e.g., architectural safety mechanism, safety design feature, safety implementation technique). Safety constraints include mechanisms such as failsafe designs, safeguards concerning electricity and the handling of toxic chemicals, the mandatory placement of warning signs, warnings to personnel, self-monitoring, exception handling, and reconfigurations (*30, 31*). In the stage of safety requirement elicitation, safety constraints arise from "the hazards that the safety control system is intended to prevent" (*32*). In the system design and implementation phase, safety constraints are further "divided into sub-requirements which are then allocated to subsystems or components" (*32*). For example, a safety constraint in subways may be "the cabin doors should stay closed when the train is moving", which should be enforced during train operation (*32*). Passengers could fall out of the train and become injured or even killed if this constraint is violated (*32*).

In a given system, the system constraints are enforced by the controllers through behaviors and interactions among the components (*33*). Each controller both enforces constraints and is enforced by constraints. In systems theory, the behaviors of each subsystem at the lower level are controlled by the constraints of the system at a higher level (*9*). Therefore, safety constraints can be classified into two types: system safety constraints and control safety constraints. System safety constraints include subsystem (or component) safety constraints and interaction safety constraints. Control safety constraints include safety control component constraints and control process safety constraints. From the perspective of systems thinking, the safety constraints of a system (or control system) can be viewed as safe pipelines or paths to maintain the directional flow of energy, materials and information in the system. Abnormal flow of energy, materials and information occurs if the safety constraints of the system and control are not enforced, which can result in unsafe behavior and conditions in the system. The path and direction of the flow of energy, materials and information in the system can be controlled by safety constraints that restrict the system structure functions and the freedom of system behavior. Emergent properties, which are related to the behavior of the system components, are controlled or enforced by a set of constraints (control laws) (*9*). External disturbances and internal disturbances are handled by the control system, which enforces the safety constraints of the system. Internal disturbances include component failures and dysfunctional interactions.

*Safety control loop and safety control structure*

The aim of the control loop (or feedback loop) of a system is to satisfy constraints on system development and system operation to react to changes in the environment (*9, 34*). Fig. 1A shows a typical control loop composed of a controller, actuator, sensor, and controlled system (or process, interaction, object). In this control loop, the actuator takes command signals as control actions from the controller, whereas the sensor feeds back the values of the controlled process to the controller (*33*). A simple control loop is composed of a controller and controlled system (or controlled process, interaction, loop, etc.). In the simple control loop, a controller, which adapts its control actions according to the process model, performs actions to influence a process or system, and then a sensor provides feedback to the controller (*35*). Complex and emergent patterns in working processes are controlled by hierarchies of controlling units (*36*). The system consists of the processes that need to be controlled, controllers that have models representing the process and a control algorithm, sensors that provide data on the states of processes, and actuators that exert control on the processes (*36*). Controllers can be stacked upon one another; the agent controlled by a higher hierarchy is essentially seen as a process that can be controlled by the same principles (*36*). The controlling unit determines the set-point at which level the controlled process should run (*36*).

Control theory provides useful insights into the requirements of safety management (*37*). Safety, which is managed by a set of control structures embedded in sociotechnical systems, can be viewed as a control problem (*9*). Safety management is thus often defined as "a continuous control task to impose the constraints necessary to limit system behavior to safe changes and adaptations" (*9*). As one of three basic concepts of STAMP, a safety control structure is based on safety control loops that have similar components as controllers, actuators, sensors and controlled processes (*9*). The control mechanisms in sociotechnical systems are modeled in a hierarchical control structure that can be created for the development and operations of complex systems (*33, 38*).

In sociotechnical systems, the human-related controllers in social systems interact with the automation-related controllers in control loops that include multiple human-related and automation-related controllers (*33*). A model of a system is used by the controller to ascertain the system's state and to enforce safety constraints on the controlled process or system (*33*). At the micro level of the system, an automated controller holds a logical model of a system, while the human-related controllers hold a mental model of the system. A safety controller, which has goals that affect the state of the system, is part of the control loop and is used to keep the system safe (*33*). In each control loop, hazardous behaviors at each level of the sociotechnical control structure result from inadequate enforcement of safety constraints on the process controlled at the level below (*9, 24*).

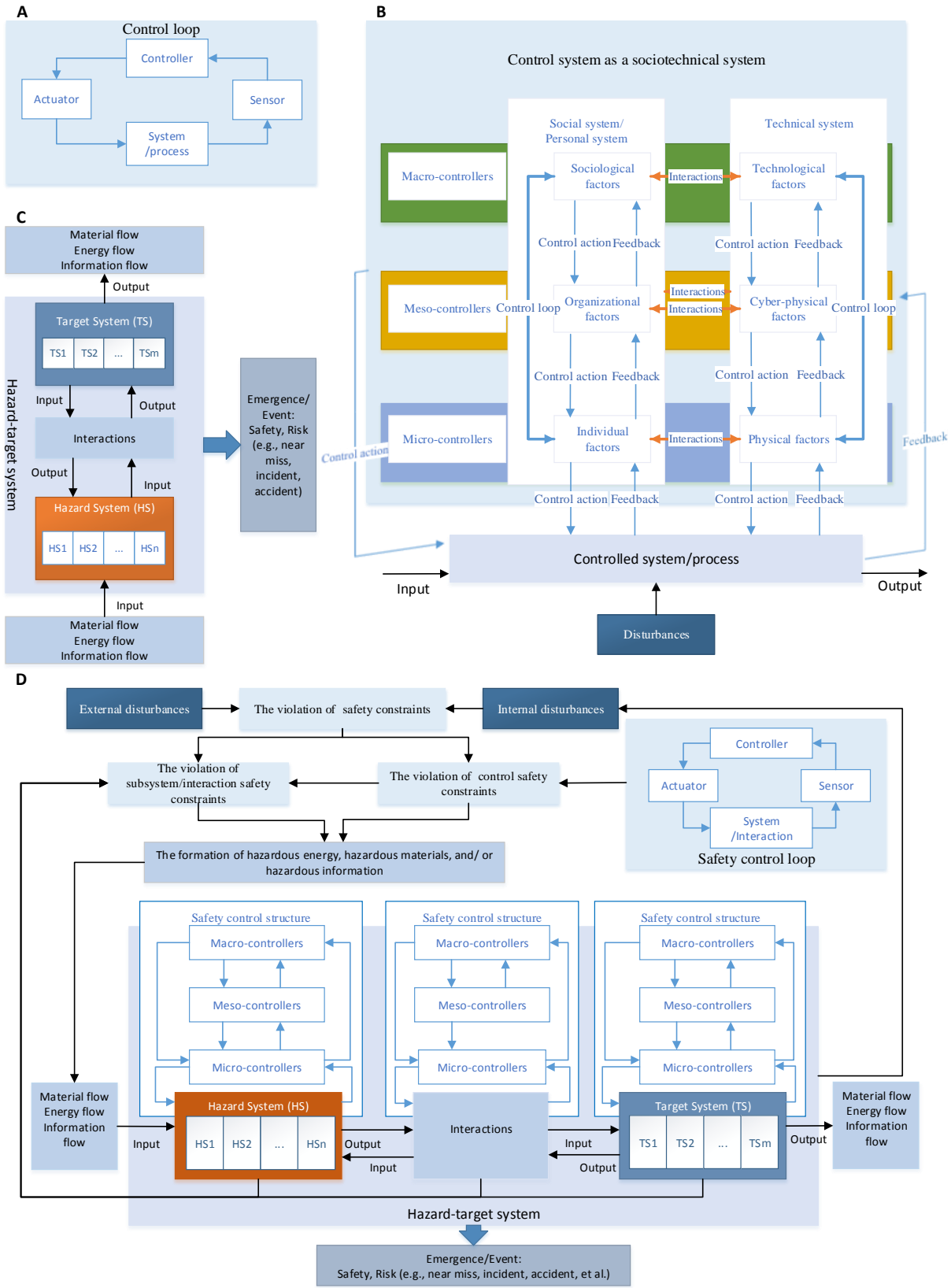

**Fig. 1 Concepts and general form of Interaction Theory of Hazard-Target System.** (**A**) A typical control loop. (**B**) Three hierarchical levels of controllers in the safety control structure of sociotechnical system. (**C**) Interactions in a hazard-target system. (**D**) General framework of Interaction Theory of Hazard-Target System

In STAMP, safety constraints in the system are enforced by control loops or the control system. In addition to physical and managerial controls, all behaviors are influenced and at least partially "controlled" by the social and organizational context in which the behavior occurs (*9*). The social aspect includes "human, economic, political, organizational and other interactions in the system" (*21*). Control is provided indirectly by social factors such as policies, procedures, and cultures (*29*). Fig. 1B shows three hierarchical levels of controllers in the control structure of the sociotechnical system. Normally, the control system has a three-level hierarchical architecture: (1) micro-level controllers that include individual factors in the social system and physical factors in the technical system; (2) meso-level controllers that include organizational factors in the social system and cyber-physical factors in the technical system; and (3) macro-level controllers that include sociological factors in the social system and technological factors in the technical system. Violations of system constraints are caused by failure of the control system. The control action path may contain mechanisms by which the controller acts upon a controlled process (referred to as actuators) and mechanisms by which the controller senses feedback from a controlled process (referred to as sensors). Each entity has control and authority over the entities immediately below, and each entity is likewise subject to control and authority from the entities immediately above.

Take work system as example. Work system can be viewed as a sociotechnical system composed of a social subsystem (or a personal subsystem) and a technical subsystem (*24, 25*). Therefore, work system is a sociotechnical ecosystem of individuals, organizations, sociology, information and technology. Safety is the outcome of interactions in a work system in which the subsystems do not bear all the responsibility of keeping the entire system and its workers safe (*24, 25*). Three hierarchical levels of safety control loops in the work safety control system is shown in fig. S1. In a work safety control system, individual factors (e.g., operators, workers, managers, specialists) as controllers at the micro-level are influenced by the organizational factors (e.g., company, governmental administration, agents) at the meso-level and sociological factors (e.g., culture, law, regulation, economics) at the macro-level. Physical factors at the micro-level (e.g., device, tools, instruction) are influenced by cyber-physical factors (e.g., cyber-physical system design, manufacture, operation, maintenance and development) at the meso-level and technical factors (e.g., informational, communication, computer, and internet technology) at the macro-level. For example, the behavior of workers is influenced by their physical and psychological conditions, their organizations and their social environment. Therefore, the behavior of workers is influenced by their organizational context and social context. Physical control equipment, such as equipment and facilities, is affected by the development of the physical information system and other technical factors. Each factor is explained in reference to other factors (i.e., the technical system is explained considering the social system, and vice versa).

*Interactions in a hazard-target system*

The hazard-target system is a new terminology that was first proposed as the object of safety science based on the Model Analysis of Dysfunctions of the System and Hazard-Barrier-Target Model (*39*). Hazards are often defined as "an inherent capability (energy, material/substance property, activity) to harm or damage due to an effect on something exposed that is valued" (*39, 40*). A hazard-target system can be defined as a "system or a system of systems that is composed of hazard systems and target systems where there are many interactions between the two components" (*39, 40*). It is composed of hazard source systems (or hazard systems) and target systems (*39*). A target system is a system composed of the targets that we want to protect from damage of the hazard system (*39*). It includes employee safety and health, the public, the environment and so on. A hazard system is a system that has hazardous energies, hazardous materials or hazardous information that can damage the target systems (*39*). There is no hazard without a target in a hazard-target system, which can be viewed as a dyad.

Both "safety" and "risk" can be viewed as the results of interactions among the components of hazard-target systems (*38, 39*). Fig. 1C shows the interactions among the components of the hazard-target system. Many target systems and hazard systems exist in hazard-target systems. The target system includes the internal and external properties and the environment of the work system that can be damaged by hazard systems. The hazard system, which is composed of hazardous energy, hazardous materials and information, can be viewed as a system that can damage a target system. Interactions in hazard-target systems include the interactions among hazard systems, the interactions among target systems, and the interactions among hazard systems and target systems. From the perspective of systems thinking, all kinds of adverse outcomes in a hazard-target system can be viewed as passive results of the interactions between hazard systems and target systems. In contrast, safety is a positive result of the interactions among components of a hazard-target system. Risks such as accidents, incidents, and near misses can be viewed as adverse events, while safety can be considered a dynamic event (*41, 42*). Therefore, safety (or risk) can be viewed as either a dynamic event in simple linear systems or as the emergence of systems in complex nonlinear systems (*20*).

*Systematic accident mechanism of the Interaction Theory of Hazard-Target System*

In RMF and STAMP models, safety is viewed as a control problem (*9*). Fig. 1D shows the general framework of ITHTS. The hazard-target system and its interactions are enforced by safety control systems (or safety control structures). Safety control structures have three hierarchical levels: micro-level, meso-level and macro-level. In a hazard-target system, hazard systems arise from the violation of hazard-target system safety constraints caused by external disturbances and internal disturbances. External disturbances include environmental changes, noise, and natural factors. Internal disturbances arise from the failure of the safety control loop, dysfunctional interactions between subsystems (or components) of the hazard-target system and component failure. The safety constraints of hazard-target systems include subsystem (or process) safety constraints, interaction safety constraints and control safety constraints. Subsystem safety constraints and interaction safety constraints are enforced by safety control loops. Control safety constraints are enforced to keep safety control loops effective. Hazard systems can be formed if hazardous materials (or energy, information) accumulate, transform and/or transfer because of the failure of safety control loops. Therefore, violations of control safety constraints can lead to violation of system (or process, interaction) safety constraints, which results in the formation of hazardous energy, hazardous materials, and/or hazardous information. Both safety and risk can be viewed as emergent properties of hazard-target systems (*39*). Safety can be seen as a positive emergence in hazard-target systems that results in favourable consequences or outcomes (e.g., occupational safety and health, the success of tasks, and production safety). Risks are viewed as a negative emergence of hazard-target systems that cause adverse consequences such as losses, injuries, occupational diseases, environmental pollution, and the failure of tasks.

The target system is at risk when safety constraints in a hazard-target system are violated, while the target system is safe when the safety constraints are all enforced by safety control loops. The target system is in a state of "near miss" when the subsystem safety constraints are partially violated. If the subsystem safety constraints are completely violated, the system is in a state of "incident". If the system safety constraints are completely violated and the target system has interacted with the hazard system, exceeding the bearing capacity of the target system, then the target system's function will fail, and people could be injured. Therefore, risks include near misses, incidents, accidents, disasters, etc.

**Case study**

Safety models cannot provide specific solutions to prevent accidents because they are always "general in purpose rather than application-specific as methods" (*43, 44*). On the other hand, accident analysis methods provide the necessary information to analyse accidents in a specific setting (*43*). Current accident analysis methods still have several problems to solve: (1) they do not address external and environmental aspects of the work domain, such as regulatory and economic influences on safety (*17*); (2) there is not enough support in these methods to analyse interactions across system levels; (3) few methods describe the details of their reliability and validity; and (4) most methods prove to be weak at representing the environment and context of the work system and do not address wider environmental issues (e.g., the role of political, legislative or regulatory factors) in shaping the overall functioning and mode of operation of work system (*17*). Current accident analysis methods have not provided a complete understanding of how dynamic, complex system behaviour contributes to an accident (*9*). A combination of model-method pairs is needed to provide a better and more reliable platform for accident analysis. Therefore, we proposed a new accident analysis method named Systemic Accident Analysis Method of Work System (SAAMWS) that is based on ITHTS and Systems Theoretical Process Analysis (STPA) (see "Systemic Accident Analysis Method of Work System" in Method).

Does this new theory explain the phenomena in safety practice? Could the new accident analysis method be applied to major accidents? To demonstrate the viability of ITHTS and the applicability of the new accident analysis method, we select the Tianjin Port fire and explosion accident in China as a case study. This major accident is "not only a representative major accident in the hazardous chemicals storage and transportation system in China, but also a rare occurrence all over the world" (*45*). It is an emergent property of a complex sociotechnical system with many stakeholders and has been analysed by many accident analysis methods (e.g., Bowtie, Accimap, STAMP, HFACS) in the literature (*45, 46, 47, 48*). In this case study, we describe the Tianjin Port fire and explosion accident in China and then analyse its accident causation factors by SAAMWS.

**Description of the Tianjin Port fire and explosion accident in China**

On August 12, 2015, the Tianjin Port fire and explosion accident occurred in China and caused serious casualties and property losses (the losses included 165 deaths, 8 missing, 789 injuries, 304 destroyed buildings, 12428 destroyed cars, and 7533 destroyed containers). At 22:51:46 on August 12, 2015, a fire of hazardous chemical containers (nitrocellulose containers) broke out in the delivery zone of the Ruihai Company hazardous goods warehouse located in the Binhai New Area of Tianjin City, China (*46*). At 23:34:06, the first explosion (its magnitude equivalent to 15 tons of TNT) occurred, and a second and more violent explosion (its magnitude equivalent to 430 tons of TNT) occurred at 23:34:37. There were six large fires and dozens of small fires at the scene of the accident, and the visible flames in these areas were not completely extinguished until 16:40 on August 14.

In the accident investigation report of the State Council of China in 2016, the nitrocellulose in the containers in the south of the dangerous goods warehouse of Ruihai Company was found to be partially dry due to the loss of wetting agent (*49*). Under high temperatures (weather), the nitrocellulose decomposed and released heat rapidly, and the accumulated heat caused spontaneous ignition. Containers of flammable chemicals near the nitrocellulose containers were ignited in succession and decomposed rapidly because of the violations of the required safety distances and the mixed storage of different types of hazardous chemicals. Containers of explosive chemicals near the hazardous chemical containers then ignited and exploded, which caused the dangerous chemicals in adjacent containers to burn and explode for a long time because of the excessive storage of hazardous goods by Ruihai Company.

**Application of SAAMWS**

*Identify the risks in the work system and define the hazard-target system*

The accident report from the Tianjin Port fire and explosion indicated that the major accident (or disaster) was related to three types of risks (or accidents, dangers): the spontaneous combustion of nitrocellulose container (near miss), the ignition of flammable chemicals containers (incident), and the explosion of hazardous goods containers (accident). Therefore, we analyse four risks (or dangers) in Ruihai Company: the spontaneous combustion of nitrocellulose (near miss), the fire incident of hazardous chemical containers (incident), the explosion accident of hazardous chemical containers (accident) and the fire and explosion accident in Tianjin Port (major accident). Fig. 2 shows the interactions in this major accident. The target system includes the persons, the values and the environment in the interior and external areas of Ruihai Company that can be influenced by the hazard systems. The persons and properties near the hazard system can be viewed as the target system (denoted by $TS$) that includes $TS1$ (*target system 1*, the persons and properties of Ruihai Company) and $TS2$ (*target system 2*, the persons, properties and the environment near the hazardous goods yard outside of Ruihai Company).

The hazardous goods yard of Ruihai Company can be viewed as the hazard system (denoted by $HS$) because it is the hazardous source of the fire and explosion accidents. The hazard system ($HS$) includes $HS1$ (*hazard system 1*, the containers of nitrocellulose), $HS2$ (*hazard system 2*, the containers of flammable chemicals near the containers of nitrocellulose) and $HS3$ (*hazard system 3*, the containers of explosive chemicals, e.g., ammonium nitrate, near the containers of flammable chemicals). Therefore, the spontaneous combustion of nitrocellulose containers (denoted by $R_1$) arises from the violated subsystem safety constraints of $HS1$. The fire incident of hazardous chemical containers (denoted by $R_2$) is the result of the dysfunctional interactions between $HS1$ and $HS2$. The explosive accident of hazardous chemical containers (denoted by $R_3$) is the adverse result of interactions between $HS$ and $TS1$. The Tianjin Port fire and explosion accident (denoted by $R_4$) is the adverse result of interactions between $HS$ and $TS$. Therefore, the hazard-target system can be viewed as being composed of Ruihai Company and its surrounding residential areas and enterprises.

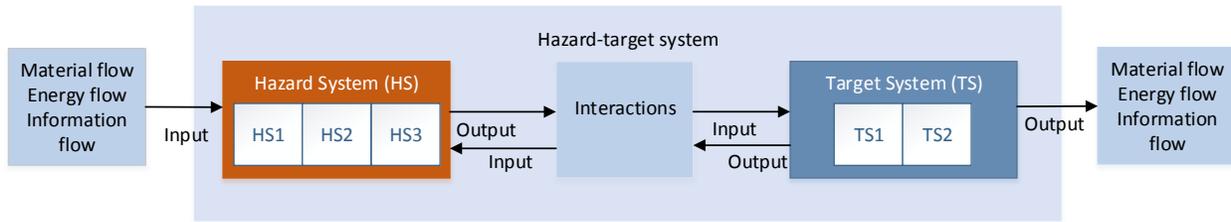

Fig. 2 The interactions in the Tianjin Port fire and explosion accident

*Model the safety control structure of controlled systems by the STPA method*

The controlled systems (or processes) include hazard systems ($HS1$, $HS2$, $HS3$, $HS$), target systems ($TS1$, $TS2$, $TS$) and the interactions among these systems. The safety constraints of the controlled systems must be enforced by the control system (or control structure, control loops). The work safety control system has three levels of safety control structures: the micro-level controllers, the meso-controllers, and the macro-controllers. Fig. 3 shows three hierarchical levels of the safety control structure of the controlled systems. The micro-controllers in the social system include humans from companies (such as front-line workers, specialists, managers in Ruihai Company, safety assessment personnel in intermediary agencies) and people in government (e.g., reviewers and inspectors). The micro-controllers in the technical systems include packages of hazardous goods, safety monitoring devices, and PPE. Organizational factors, such as the meso-controllers in the social system, refer to the Tianjin Government and its bodies, Ruihai Company, intermediary agencies, social organizations and media. Cyber-physical factors, such as the meso-controllers in the technical system, refer to safety design, safety maintenance, safety operation and safety development. The sociological factors, such as the macro-controllers in the social system, include culture, law, regulation, policy, and economics. The technological factors as the macro-controllers in the technical systems, include communication, computer, and packaging.

In the micro-level analysis, the micro-controllers include frontline workers, operators, specialists, managers, and supervisors. For example, frontline workers were responsible for the handling, loading, unloading and storage of hazardous goods. In the meso-level analysis, organizational factors include government safety administration (e.g., government management and coordination, regulatory agents' supervision and approval), intermediary agents' safety services (e.g., design of renovation projects, safety assessment, and safety reviews), social organizations' supervision (e.g., media reports on the illegal activities of Ruihai Company), and Ruihai Company's safety management (e.g., supplier management, process and personal safety management). For example, the Tianjin government is responsible for (1) the coordination and management of hazardous chemicals in Tianjin Port and (2) guidance and supervision of subordinate governments. Tianjin Customs, as one of the regulatory agents, has responsibility for the approval of individual inspection items and regular supervision stipulated in regulations. In the macro-level analysis, sociological factors include law (e.g., national law, government regulation, and standards), policies (government policy, corporate policy), organizational culture (e.g., government culture, corporate safety culture), economics, and others. For example, Ruihai Company safety management is "controlled" by its safety culture. Therefore, the meso-controllers and the macro-controllers in the social system can be viewed from the organizational context and sociological context, respectively, of the individual factors.

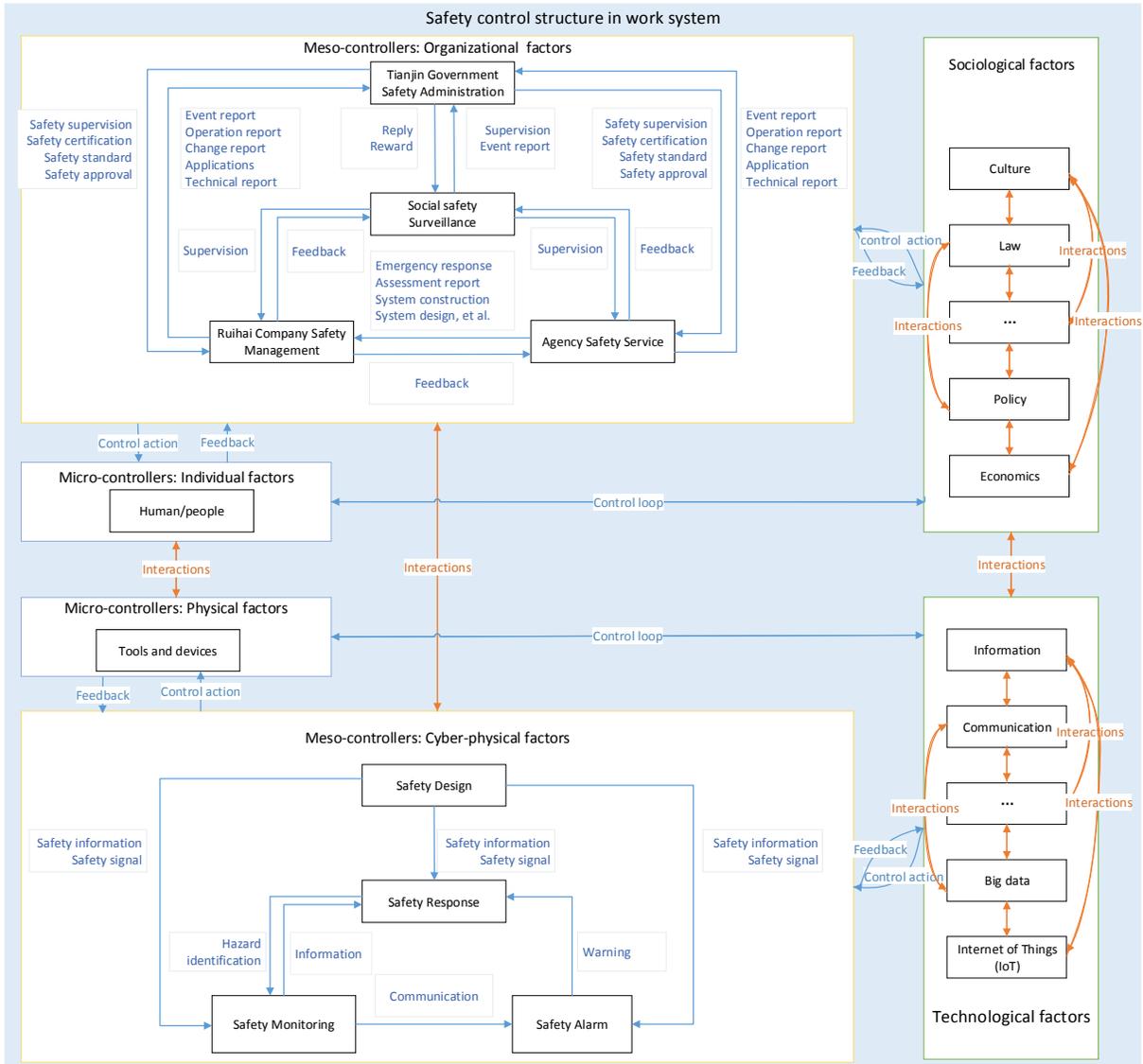

Fig. 3 Three hierarchical levels of the safety control structure of the controlled systems

*Identify the violations of safety constraints and convert them into adverse events*

The violations of safety constraints in Tianjin Port fires and explosion accidents can be analysed by the STPA method. However, the analyses are not repeated in this paper because they have been reported in detail in the literature (*47, 50, 51*). The safety constraints in hazard-target systems can be classified into three types: subsystem safety constraints, interaction safety constraints and control safety constraints. For example, the subsystem safety constraints of $HS1$ are as follows: (1) nitrocellulose must be damped with wetting agents (denoted by $SC_{1.1}$); (2) the packaging of nitrocellulose with wetting agents must be sealed well (denoted by $SC_{1.2}$); (3) nitrocellulose packaging must not be damaged (denoted by $SC_{1.3}$); and (4) the storage temperature of nitrocellulose must be lower than 35℃ (denoted by $SC_{1.4}$). The interaction of safety constraints $HS1$-$HS2$ refers to the required distance between the nitrocellulose containers and the flammable chemicals according to the legal safety distances (denoted by $SC_{1.9}$). At the micro level, the control safety constraints of $HS1$ are as follows: (1) the storage of the nitrocellulose containers must maintain good ventilation or take cooling measures to ensure an appropriate ambient temperature (denoted by $SC_{1.5}$); (2) nitrocellulose damped with wetting agents must be sealed in plastic film (denoted by $SC_{1.6}$); (3) The operation of the nitrocellulose containers must be complied with safety rules (denoted by $SC_{1.7}$); and (4) fire fighting measures must be taken quickly and effectively if the nitrocellulose containers spontaneously combust (denoted by $SC_{1.8}$).

| Violation of safety constraints in the hazard-target system at the micro-level analysis | The corresponding adverse events in the hazard-target system at the micro-level analysis |
|---|---|
| **The Hazard System (HS for short) includes HS1, HS2 and HS3**<br>**HS: the hazardous goods yard in Ruihai Company**<br>**HS1 (Hazard System 1): the containers of nitrocellulose**<br>The subsystem safety constraints of HS1<br>SC1.1: Nitrocellulose must be damped with wetting agents<br>SC1.2: The packaging of nitrocellulose with wetting agents must be sealed well<br>SC1.3: Nitrocellulose packaging must not be damaged<br>SC1.4: The storage temperature of nitrocellulose must be lower than 35℃<br>The micro-level control safety constraints of HS1<br>SC1.5: The storage of the nitrocellulose containers must maintain good ventilation or take cooling measures to ensure an appropriate ambient temperature<br>SC1.6: Nitrocellulose damped with wetting agents must be sealed in plastic film<br>SC1.7: The operation of the nitrocellulose containers must be complied with safety rules<br>SC1.8: The fire fighting measures must be taken quickly and effectively if the nitrocellulose containers spontaneously combust<br>**HS2 (Hazard system 2): the containers of inflammable chemicals nearby the containers of nitrocellulose**<br>The micro-level control safety constraints of the interactions of HS1-HS2<br>SC1.9: The distances between the containers of nitrocellulose and the inflammable chemicals must be accorded with the requirements of legal safety distances<br>The micro-level control safety constraints of HS2<br>SC1.10: If the fire of hazardous goods appears, the control measures must be taken timely and effectively<br>**HS3 (Hazard system 3): the containers of explosive chemicals (e.g. ammonium nitrate) nearby the containers of inflammable chemicals**<br>The micro-level control safety constraints of the interactions of HS1-HS2-HS3<br>SC1.11: The inflammables and explosive chemicals must not be stored together<br>The micro-level control safety constraints of HS3<br>SC1.12: The hazardous goods must not be stored in excess<br>**The Target System (TS for short) includes TS1 and TS2**<br>**TS1 (Target system 1): The persons and properties in Ruihai Company**<br>**TS2 (Target system 2): The persons, properties and the environment nearby the hazardous goods yard at the outside of Ruihai Company**<br>The micro-level control safety constraints of the TS (TS1 and TS2)<br>SC 1.13: The emergency evacuation should be taken timely and effectively before the exploration accident occurs<br>**The interactions between the HS and TS**<br>The micro-level control safety constraints of the interactions of HS-TS:<br>SC1.14: The distances between the hazardous goods yard and the neighboring residential areas must be accorded with the requirements of legal safety distances | **The Hazard System (HS for short) includes HS1, HS2 and HS3**<br>**HS: the hazardous goods yard in Ruihai Company**<br>**HS1 (Hazard system 1): the containers of nitrocellulose**<br>The subsystem safety constraints of HS1<br>E1.1: The loss of the nitrocellulose wetting agent<br>E1.2: Poor sealing of the nitrocellulose packaging<br>E1.3: Damage to the nitrocellulose packaging<br>E1.4: High environmental temperature (weather)<br>The micro-level control safety constraints of HS1<br>E1.5: No cooling measures<br>E1.6: The nitrocellulose damped with alcohol being stored in plastic bags<br>E1.7: Rough handling by workers<br>E1.8: Failure to take appropriate measures in time<br>**HS2 (Hazard system 2): the containers of inflammable chemicals nearby the containers of nitrocellulose**<br>The micro-level control safety constraints of the interactions of HS1-HS2<br>E1.9: Violation of the safety distances between the flammable chemical containers and the nitrocellulose containers<br>The micro-level control safety constraints of HS2<br>E1.10: Failing to control the fire of inflammable chemicals<br>**HS3 (Hazard system 3): the containers of explosive chemicals (e.g. ammonium nitrate) nearby the containers of inflammable chemicals**<br>The micro-level control safety constraints of the interactions of HS1-HS2-HS3<br>E1.11: Mixed storage of inflammable goods and explosive chemicals<br>The micro-level control safety constraints of HS3<br>E1.12: Excessive storage of hazardous goods<br>**The Target System (TS for short) includes TS1 and TS2**<br>**TS1 (Target system 1): The persons and properties in Ruihai Company**<br>**TS2 (Target system 2): The persons, properties and the environment nearby the hazardous goods yard at the outside of Ruihai Company**<br>The micro-level control safety constraints of the TS (TS1 and TS2)<br>E1.13: Delayed or ineffective emergency evacuation<br>**The interactions between the HS and TS**<br>The micro-level control safety constraints of the interactions of HS-TS:<br>E1.14: Violating the safety distances between the hazardous goods yard and the neighboring residential areas |

Fig. 4 The violations of safety constraints and adverse events in the micro-level analysis

Fig. 4 shows the violations of safety constraints in the hazard-target system and the corresponding adverse events in the micro-level analysis. Adverse events at all hierarchical levels can be viewed as the result of violating safety constraints in the hazard-target system. For simplicity, it is assumed that only one adverse event corresponds to each safety constraint that is violated in the system. In other words, each safety constraint (denoted by $SC$) that is violated in the hazard-target system has only one corresponding adverse event (denoted by $E$) in this case study. Therefore, the corresponding adverse events resulting from the above subsystem safety constraint violations in $HS1$ are as follows: (1) the loss of the nitrocellulose wetting agent (denoted by $E_{1.1}$); (2) poor sealing of the nitrocellulose packaging (denoted by $E_{1.2}$); (3) damage to the nitrocellulose packaging (denoted by $E_{1.3}$); and (4) high environmental temperature (weather) (denoted by $E_{1.4}$). Other violated safety constraints and the corresponding adverse events can be obtained in the same way. In terms of the meso-level analysis, the violations of safety requirements of Ruihai Company safety management are shown in table S1. The violations of safety constraints in the hazard-target system and the corresponding adverse events in the meso-level analysis and the macro-level analysis are presented in fig. 2S and fig. 3S, respectively.

*Draw the event flow diagram of the risks at different levels*

Fig. 5 shows an event flow diagram of the Tianjin Port fire and explosion accident at the micro-level and meso-level. In the micro-level analysis, the near miss in Ruihai Company is the spontaneous combustion of nitrocellulose (denoted by $R_1$), which was caused by the loss of the nitrocellulose wetting agent (denoted by $E_{1.1}$) and high environmental temperature (denoted by $E_{1.4}$). The loss of the nitrocellulose wetting agent ($E_{1.1}$) refers to two adverse events: poor sealing of the nitrocellulose packaging ($E_{1.2}$) caused by the nitrocellulose damped with alcohol being stored in plastic bags ($E_{1.6}$), and damage to the nitrocellulose packaging ($E_{1.3}$) caused by rough handling by workers ($E_{1.7}$). The fire incident of hazardous chemical containers in Ruihai Company (denoted by $R_2$) was caused by three factors: the spontaneous combustion of nitrocellulose ($R_1$), failure to take appropriate measures in time ($E_{1.8}$) and violation of the safety distances between the flammable chemical containers and the nitrocellulose containers ($E_{1.9}$). The explosion accident of hazardous chemical containers in Ruihai Company (denoted by $R_3$) was caused by three factors: the fire incident of hazardous chemicals containers ($R_2$), failing to control the fire of flammable chemicals ($E_{1.10}$) and mixed storage of inflammable goods and explosive chemicals ($E_{1.11}$). The fire and explosion accident of hazardous chemical containers in Ruihai Company (denoted by $R_4$) was caused by four factors: the fire incident of hazardous chemicals containers ($R_3$), excessive storage of hazardous goods ($E_{1.12}$), delayed or ineffective emergency evacuation ($E_{1.13}$) and violating the safety distances between the hazardous goods and the neighbouring residential areas ($E_{1.14}$).

In the meso-level analysis, adverse event $E_{1.5}$ was influenced by four contributors: lack of safety training and education ($E_{2.2}$), failure to implement safe operation rules ($E_{2.3}$), careless review without on-sit inspection ($E_{2.16}$), irresponsible on-site inspection and examination ($E_{2.17}$). Adverse event $E_{1.6}$ was influenced by two contributors: poor supplier management of hazardous goods ($E_{2.1}$) and improper packaging design of nitrocellulose ($E_{2.28}$). The adverse event $E_{1.7}$ was caused by three contributors: lack of safety training and education ($E_{2.2}$), failure to implement safe operation rules ($E_{2.3}$) and lack of on-site supervision ($E_{2.4}$). Other adverse events at the micro-level can be represented by the corresponding adverse events at the meso-level analysis in the same way.

The adverse events at the macro-level impact the adverse events at both the meso-level and micro-level. For example, adverse event $E_{3.1}$ (bad government culture) at the macro-level can impact the following three adverse events at the meso-level: $E_{2.15}$ (approval through unconventional procedure), $E_{2.16}$ (careless review without on-sit inspection), and $E_{2.17}$ (irresponsible on-site inspection and examination). Adverse event $E_{3.3}$ (managers and workers were accustomed to violating the safety rules to pursue efficiency and other benefits) at the macro-level impacts the following three adverse events at the meso-level: $E_{2.2}$ (lack of safety training and education), $E_{2.3}$ (failure to implement safe operation rules), $E_{2.4}$ (lack of on-site supervision), $E_{2.5}$ (lack of knowledge about what substances are on fire), $E_{2.6}$ (lack of emergency evacuation training of workers and surrounding residents), $E_{2.7}$ (blocked fire passages) and $E_{2.28}$ (improper packaging design of nitrocellulose). It can also influence the following adverse events at the micro-level: $E_{1.7}$ (rough handling by workers), $E_{1.11}$ (mixed storage of flammable goods and explosive chemicals) and $E_{1.12}$ (excessive storage of hazardous goods). The adverse events at the macro-level and the corresponding relationship to the adverse events at the meso-level and micro-level are shown clearly in table S2.

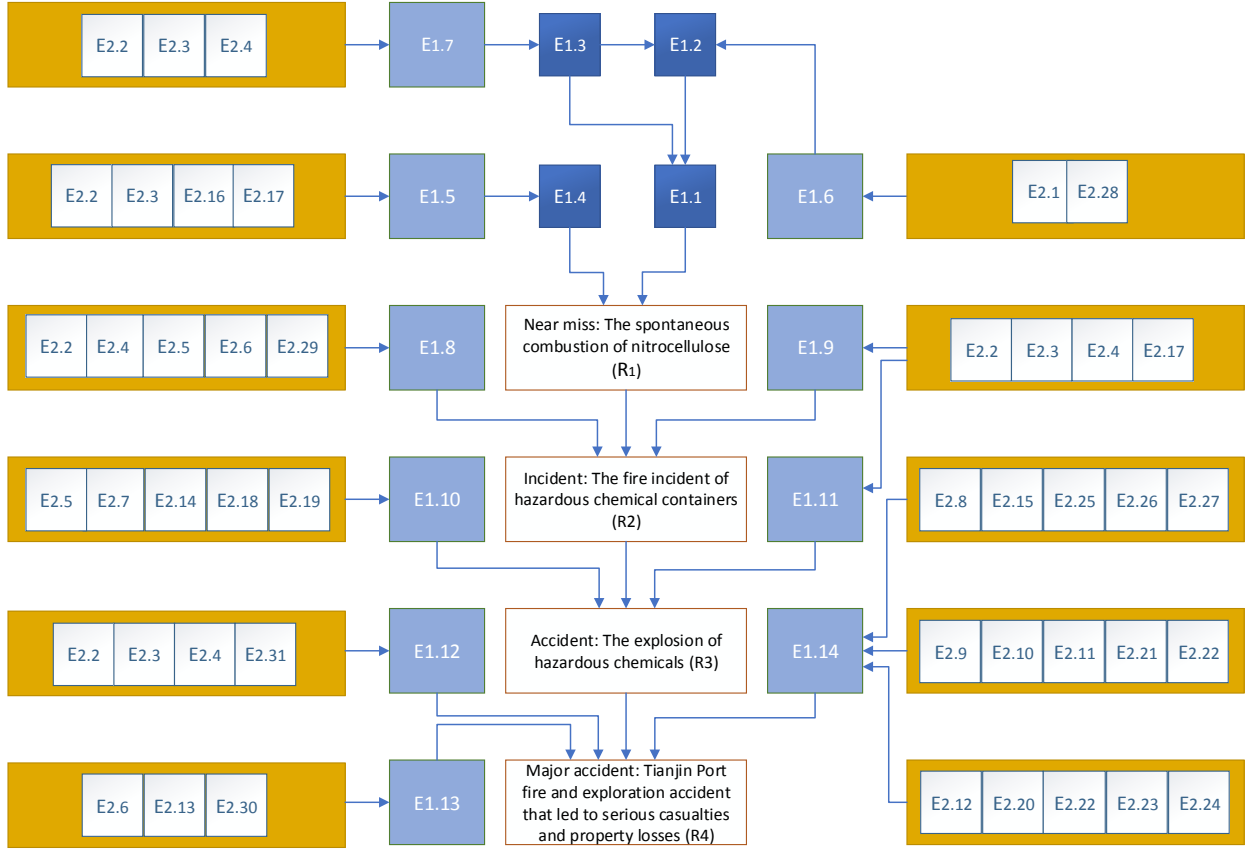

Fig. 5 The event flow diagram at the micro-level and meso-level

## Recommendations

SSAMWS provides an exhaustive description of the safety requirements and responsibilities of many stakeholders, such as companies, governments, regulatory bodies, intermediary agencies, social organizations and media. For instance, several persons in the Tianjin government and some of the regulators wilfully failed to obey the law. They allowed Ruihai Company to operate without the necessary licences and ignored every violation of law pertaining to the control of risks associated with handling large quantities of dangerous goods *(46)*. The list of recommendations is shown in table S3.

## Discussion

To solve the problems of existing accident models, we propose the Interaction Theory of Hazard-Target System (ITHTS), which incorporates human, organizational and technological characteristics in the same framework and considers contributory factors such as legislative, regulatory, and cultural factors across all levels of sociotechnical system. The applications of this theory are not limited to specific systems (e.g., simple linear system, complex linear system, dynamic nonlinear system). Some accident models (e.g., domino model, Hazard-Barrier-Target Model, Swiss Cheese Model, RMF) can be regarded as simplified or refined specific forms of this theory for different systems. Therefore, this work takes the first step towards a unified theory of accident causation in safety science.

To address the issues that current accident analysis methods face, we proposed a new systemic accident analysis method called Systemic Accident Analysis Method of Work System based on ITHTS and STPA. We chose the Tianjin Port fire and explosion accident in China as a case study to demonstrate the viability of the ITHTS and the applicability of this new accident analysis method. Our results indicate that the "root causes" of major accidents are a combination of sociological factors (e.g., culture, legislative, economics and policy) and technological factors (e.g., communication, information) that influences on the organisational factors and individual factors contributed to major accidents.

**Methods**

**Systems-Theoretic Accident Model and Processes**

Systems-Theoretic Accident Model and Processes (STAMP) proposed by Leveson in 2004, is an accident model based on systems theory (*9, 52*). In STAMP, safety is viewed as an emergency of the system that arises from interactions between the components of the sociotechnical system (*9*). In systems theory, systems are viewed as hierarchical levels of controls and constraints. Each level in the hierarchy of controls imposes constraints on the level below. The control mechanisms in sociotechnical systems are modelled in a hierarchical control structure that can be created for the development and operations of complex systems (*38*). General form of the safety control structure of a sociotechnical system is presented in fig. 4S. In this model, there are two basic hierarchical control structures: one for system development on the left and another for system operation on the right (*9*). The control action path may contain two mechanisms: one by which the controller acts upon a controlled process (referred to as actuators) and another by which the controller senses feedback from a controlled process (referred to as sensors). These details are usually abstracted away during the initial creation of the control structure and are refined later during the scenario creation step. The vertical axis in a hierarchical control structure represents control and authority within the system. The vertical placement indicates the hierarchy of control from the top (high-level controllers) to the bottom (the lowest-level entities). This safety control structure changes over time.

**Systemic Accident Analysis Method of Work System**

Many safety practitioners acknowledge the limitations of traditional sequential analysis methods such as Fault Tree Analysis (FTA) and Event Tree Analysis (ETA) and are keen to apply new techniques (*53*). The systemic accident analysis methods (SAA methods) provide "a deeper understanding of how dynamic, complex system behavior contributed to the event", while the traditional methods are "suitable for describing what happened in an accident" (*53*). They provide a comprehensive understanding of accident causation and more effective safety recommendations than traditional methods (*53*). However, the large gap between academic developments and the application of SAA methods needs to be closed (*53*). A combination of model-method pairs could provide a better and more reliable platform for accident analysis. This approach is helpful for us to better understand the complex organizational, sociological, and technological factors contributing to accidents. To address these issues, a novel method named systematic accident analysis method of work system (SAAMWS for short) is proposed based on ITHTS and STPA. The general procedure of SAAMWS can be divided into the following five steps:

*Step 1: Identify the risks of work system and define the hazard-target system*

The first step is to identify the risks of work system that need to be analysed. The risks of the work system often refer to adverse events such as fire, falling, explosion, and so on. Risks can be viewed as the results of interactions between hazard systems and target systems. Therefore, these interactions must be systematically identified. The definition of the hazard-target system should be considered the purpose and the accident analysts. Hazard systems in work system should be identified in terms of the risk that needs to be analysed and the target systems. In addition, based on the requirements and purposes of accident analysis, the interactions between hazard systems and target systems in work system should be stated and explained in this step.

Both safety and risk can be viewed as the result of dysfunctional interactions among components of the hazard-target system (*39*). Each can also be considered an emergent property that arises from interactions among the components (*38, 41*). For example, an explosion accident can be viewed as the result of interactions between hazard systems and target systems in a hazard-target system in which the storage system for explosive substances is viewed as a hazard system and a chemical factory can be viewed as a target system. Therefore, hazard systems and their interactions in the work system must be identified first to assess the risks of the work system. For simplicity, each risk is assumed to correspond to a hazard-target system.

*Step 2: Model the safety control structure of controlled system*

The control structure of the controlled system is constructed in this step. In the ITHTS, hazard systems, target systems and the interactions between them must be controlled by safety control systems. Different safety analysts may construct different safety control structures for the same controlled system. For example, the accident analysis contents of internal accident investigators may be limited to safety management and safety culture, while accident investigators in government departments may construct a safety control structure that includes the safety supervision of the government. Controlled systems include hazard systems, target systems and their interactions. The safety constraints of controlled systems must be enforced by the control system (or control structure, control loops).

*Step 3: Identify the violation of safety constraints and convert them into corresponding adverse events*

In this step, the violations of safety constraints are identified by the STPA method and then classified with respect to the three hierarchical levels of the sociotechnical control system. The violations of safety constraints are analysed and classified by means of accident analysts. The subsystem safety constraints and interaction safety constraints are enforced by safety control loops. The failure of control loops is due to violations of control safety constraints. Adverse events at all hierarchical levels can be viewed as the result of violating safety constraints in hazard-target systems. For simplicity, it is assumed that there is only one adverse event corresponding to each safety constraint that is violated in a system. In other words, each safety constraint that is violated in a hazard-target system has only one corresponding adverse event.

*Step 4: Draw the event flow diagram of the accident at different hierarchical levels*

In the sequence accident model, accident causes are often divided into three types: immediate causes, indirect causes, and root causes. Adverse events at the micro-level can be viewed as immediate causes, adverse events at the meso-level can be viewed as indirect causes, and adverse events at the macro-level can be viewed as root causes. Basic events at all hierarchical levels can be viewed as the result of the violation of safety constraints in a hazard-target system. Therefore, the sequence of the event flow is as follows: violations of control safety constraints at the macro-level, violations of control safety constraints at the meso-level, violations of control safety constraints at the micro-level, and violations of subsystem (or interaction) safety constraints of the controlled system (or process).

*Step 5: Recommendation*

The recommendations for improvement measures must be listed according to the results of accident analysis. The list of suggestions should include (1) national laws, policies, regulations and standards; (2) government and regulation bodies, including government culture, policy and regulation and safety management; (3) companies, including corporate culture, policy, and safety management; (4) intermediary agencies; (5) social organizations and media; and (6) technical aspects.

**Acknowledgments:** This study was supported by National Natural Science Foundation of China (52074066, 52004055), the National Key Program of Research and Development, Ministry of Science and Technology (2018YFC0808406), PR China, The Key Science and Technology Research Project of Jilin Institute of Chemical Technology (201909), PR China, The Science and Technology Research Project of Jilin Institute of Chemical Technology (2017052 ), PR China, Jilin Province Fundamental Research Special Fund by Central Government (202002009JC), PR China.

**Author contributions:** Conceptualization and the proposer of the ITHTS and SAAWS: J.Ge. Methodology: J. Ge. Design: J. Ge, K.L. Xu, X.W. Yao. Analysis of case study: J. Ge, Y.Y. Zhang, J.S. Li, C.Y. Wu, S.Y.Li, F. Yang, J.J. Zhang, Q.W. Xu, X.W. Yao, K.L. Xu. Funding acquisition: K.L. Xu, J. Ge, X.W. Yao, C.Y. Wu. Project administration: K.L. Xu, J. Ge, X.W. Yao. Writing – original draft: JG. Writing – review & editing: J. Ge, K.L. Xu, X.W. Yao.

**Competing interests:** Authors declare that they have no competing interests.

**REFERENCES AND NOTES**

1. K. Jayaraman, Bhopal disaster: Technical inquiry under way. *Nature* **313**, 89 (1985). https://doi.org/10.1038/313089b0

2. L.R. Anspaugh, R.J. Catlin, M. Goldman, The global impact of the Chernobyl reactor accident. *Science* **242**, 1513-1519 (1988).

3. R. L. Dillon, New ways to learn from the Challenger Disaster: Almost 30 years later. *2015 IEEE Aerospace Conference*, 1-8 (2015)

4. P.C. Burns, R. C. Ewing, A. Navrotsky, Nuclear fuel in a reactor accident. *Science* **335**, 1184-1188 (2021).

5. Z. Tang, Q. Huang, Y. Yang, Overhaul rules for hazardous chemicals. *Nature* **525**, 455 (2015).

6. J. Whitfield, Personal Factors in Accident Proneness. *Nature* **175**, 1112 (1955)

7. N. Karanikas, A. Roelen, The Concept towards a Standard Safety Model (STASAM v. 0). *MATEC Web of Conferences* **273**, 02001 (2019) https://doi.org/10.1051/matecconf/201927302001

8. J. Ge, et al., The main challenges of safety science. *Saf. Sci.* **118**, 119–125 (2019).

9. N.G. Leveson, Engineering a Safer World. (MIT Press, Cambridge, MA, 2012).

10. J. Rasmussen, Risk management in a dynamic society: a modelling problem. *Saf. Sci.* **27**(2-3), 183-213 (1997)

11. C. Perrow, Normal Accidents: Living with High-Risk Technologies (Basic Books, New York, 1984).

12. E. Hollnagel, FRAM: The Functional Resonance Analysis Method: Modelling Complex Socio-technical Systems. (Ashgate Publishing Ltd., Farnham, 2012)

13. P. Swuste, W. Zwaard, J. Groeneweg, F. Guldenmund, Safety professionals in the Netherlands. *Saf. Sci.* **114**, 79–88 (2019). https://doi.org/10.1016/j.ssci.2018.12.015

14. L. Alessandretti, U. Aslak, S. Lehmann, The scales of human mobility. *Nature* **587**, 402–407 (2020).

15. P. Waterson, et al., Defining the methodological challenges and opportunities for an effective science of sociotechnical systems and safety. *Ergonomics* **58**(4), 565-599 (2015)

16. A. Hulme, et al., What do applications of systems thinking accident analysis methods tell us about accident causation? A systematic review of applications between 1990 and 2018. *Saf. Sci.* **117**, 164–183 (2019)

17. T. Reiman, C. Rollenhagen, Does the concept of safety culture help or hinder systems thinking in safety?. *Accident Analysis & Prevention* **68**, 5-15 (2014)

18. J. F. Lutsko, How crystals form: A theory of nucleation pathways. *Sci. Adv.* **5**, 7399 (2019).

19. S.B. Benjamin, J.F. Wolter, System Engineering and Analysis (Tsinghua University Press, ed. 3, Beijing, 2002).

20. P. Blokland, G. Reniers, Safety Science, a Systems Thinking Perspective: From Events to Mental Models and Sustainable Safety. *Sustainability* **12**(12), 5164 (2020).

21. S. Ghazinoory, et al., Why do we need 'Problem-oriented Innovation System (PIS)'for solving macro-level societal problems? *Technol. Forecast. Soc. Change* **150,** 119749 (2020).

22. F.W. Geels, From sectoral systems of innovation to socio-technical systems: insights about dynamics and change from sociology and institutional theory. *Res. Policy* **33** (6–7), 897–920 (2004).

23. B. Fischhoff, The realities of risk-cost-benefit analysis. *Science* **350**, aaa6516 (2015).

24. T. Niskanen, K. Louhelainen, M. L. Hirvonen, A systems thinking approach of occupational safety and health applied in the micro-, meso- and macro-levels: a finnish survey. *Saf. Sci.* **82**, 212-227 (2016).

25. L. A. Murphy, M. M. Robertson, P. Carayon, The next generation of macroergonomics: Integrating safety climate. *Accident Analysis & Prevention* **68**, 16-24 (2014).

26. C.E. Siemieniuch, M. A. Sinclair, Extending systems ergonomics thinking to accommodate the socio-technical issues of systems of systems. *Applied Ergonomics* **45**(1), 85-98 (2014).

27. C. Cimini, et al., A human-in-the-loop manufacturing control architecture for the next generation of production systems. *Journal of Manufacturing Systems* **54**, 258–271 (2020).

28. L. Atzori, et al., The social internet of things (SIOT) – when social networks meet the internet of things: concept, architecture and network characterization. *Computer Networks* **56**(16), 3594-3608 (2012).

29. N. G. Leveson, Applying systems thinking to analyze and learn from events. *Saf. Sci.* **49**(1), 55-64 (2011).

30. D.G. Firesmith, Engineering Safety requirements, safety constraints, and safety-critical requirements. *Journal of Object Technology* **3**(3), 27-42. (2004)

31. H. H. Dreany, R. Roncace, A cognitive architecture safety design for safety critical systems. *Reliability Engineering & System Safety* **191**, 106555 (2019).

32. X. Han, T. Tang, J. Lv, A hierarchical verification approach to verify complex safety control systems based on STAMP. *Science of Computer Programming* **172**(MAR.1), 117-134 (2019).

33. D. Schmid, N.A. Stanton, How are laser attacks encountered in commercial aviation? A hazard analysis based on systems theory. *Saf. Sci.* **110**, 178-191 (2018).

34. S.Y. Gong, C.M. Guo, A safety verification method based on control constraints. *Missiles and Space Vehicles* **304**(6), 55-59 (2009).

35. J.M. Rising, N.G. Leveson, Systems-theoretic process analysis of space launch vehicles. *J. Space Saf. Eng.* **5**, 153–183 (2018).

36. D. Passenier, A. Sharpanskykh, R. J. De Boer, When to STAMP? A case study in aircraft ground handling services. *Procedia Engineering* **128**, 35-43 (2015).

37. M. B. Wahlström, C. Rollenhagen, Safety management – a multi-level control problem. *Saf. Sci.* **69**, 3-17 (2014).

38. N.G. Leveson, Rasmussen's legacy: A paradigm change in engineering for safety. *Applied Ergonomics* **59**, 581-591 (2017)

39. J. Ge, et al., What is the object of safety science?. *Saf. Sci.* **118**, 907-914 (2019).

40. H.J. Pasman, Risk assessment: what can it do for you? It may be a matter of to be or not to be!. *J. Appl. Packag. Res.* **8** (1), 8–14 (2016).

41. H. Slim, S. A. Nadeau, Proposal for a Predictive Performance Assessment Model in Complex Sociotechnical Systems Combining Fuzzy Logic and the Functional Resonance Analysis Method (FRAM). *American Journal of Industrial and Business Management* **9**, 1345-1375 (2019).

42. E. Hollnagel, Safety-I and Safety-II: The Past and Future of Safety Management. (Ashgate Publishing, Ltd., Farnham 2014).

43. P. Katsakiori, G. Sakellaropoulos, E. Manatakis, Towards an evaluation of accident investigation methods in terms of their alignment with accident causation models. *Saf. Sci.* **47**(7), 1007-1015 (2009).

44. M. Lehto, G. Salvendy, Models of accident causation and their application: Review and reappraisal. *Journal of Engineering and Technology Management* 8(2), 173-205 (1991)

45. C. Wu, L. Huang, A new accident causation model based on information flow and its application in Tianjin Port fire and explosion accident. *Reliability Engineering & System Safety* **182**, 73-85 (2018). https://doi.org/10.1016/j.ress.2018.10.009

46. Y. Chen, et al., An accident causation model based on safety information cognition and its application. *Reliability Engineering & System Safety* **207**, 107363 (2021).

47. Y. Zhang, L. Jing, C. Sun, Systems-Based Analysis of China-Tianjin Port Fire and Explosion: A Comparison of HFACS, AcciMap, and STAMP. *Journal of Failure Analysis and Prevention* **18**, 1386–1400 (2018).

48. G. Fu, J.H. Wang, M.W. Yang, Anatomy of Tianjin Port fire and explosion: process and causes. *Process Safety Progress* **35**(3), 216-220 (2016).

49. Q.L. Chen, M. Wood, J.S. Zhao, Case study of the Tianjin accident: Application of barrier and systems analysis to understand challenges to industry loss prevention in emerging economies, *Process Saf. Environ. Prot.* **131** 178–188 (2019)

50. P.F. Du, Resilience Theory-based Analysis of Accidents Framework: Take the Tianjin Port Explosion Accident as an Example (Huazhong University of Science and Technology, Wuhan, 2017).

51. Y. Zhang, et al. Systems approach for the safety and security of hazardous chemicals. *Maritime Policy & Management* **47**(4), 1-23 (2020).

52. W. Li, L. Zhang, W. Liang, Accident causation analysis and taxonomy (ACAT) model of complex industrial system from both system safety and control theory perspective. *Saf. Sci.* **92**, 94–103 (2017).

53. P. Underwood, P. Waterson, Systemic accident analysis: examining the gap between research and practice. *Accident Analysis & Prevention* **55**, 154-164 (2013).
20

# Supplementary Materials

**Materials and Methods**

**Supplementary Text**

**Figs. S1 to S4**

**Tables S1 to S3**

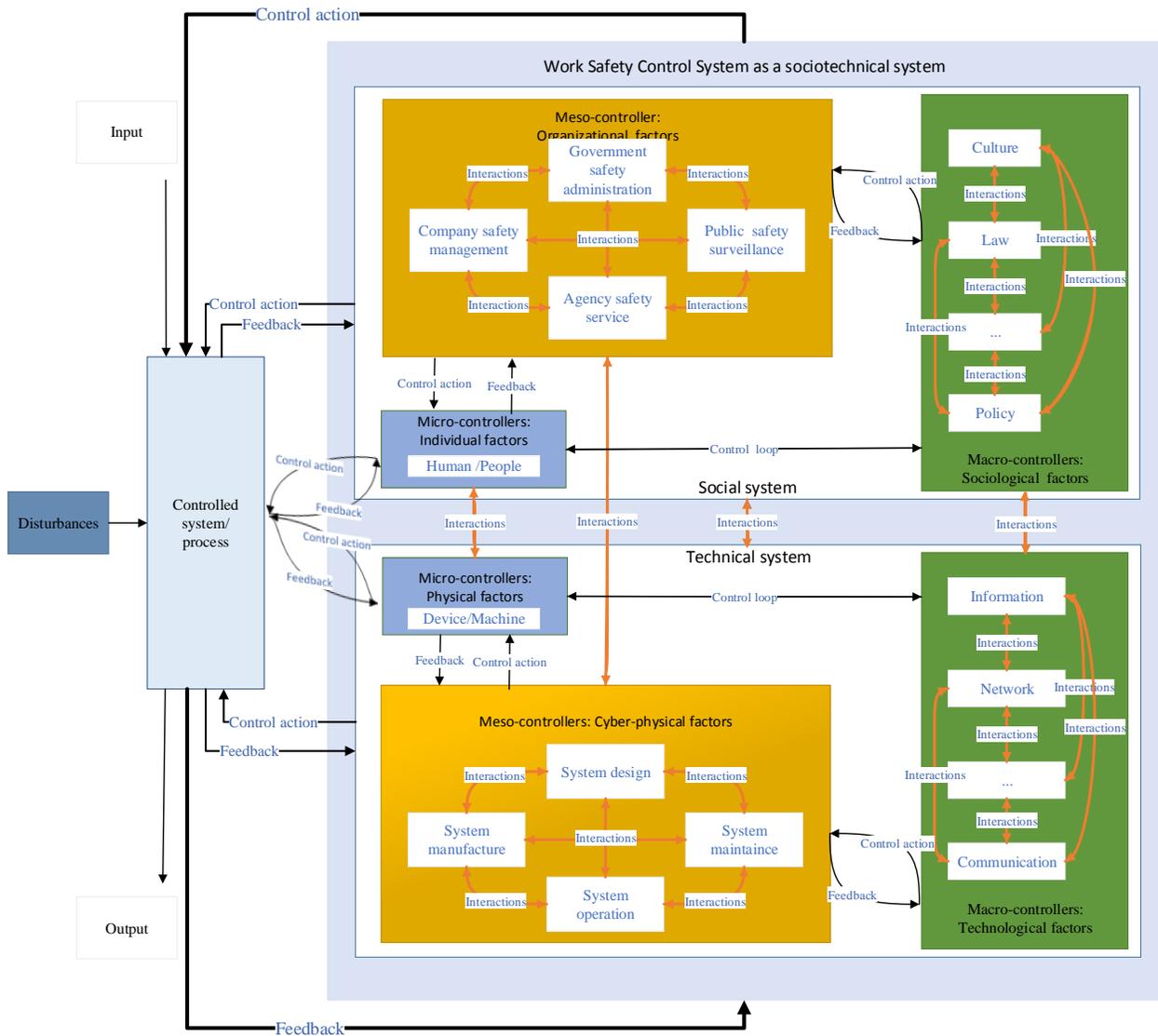

**Fig. S1. General form of safety control structure in the work safety control system**

| Violation of safety constraints in the hazard-target system at the meso-level analysis | The corresponding adverse events at the meso-level analysis |
|---|---|
| **Organizational factors in Social System**<br>**Ruihai Company Safety Management**<br>Supplier Management<br>SC2.1: The supplier management must be filled with the safety requirements<br>Process and Personal Safety Management<br>SC2.2: The workers must be trained and educated well with safety skills and safety knowledge<br>SC2.3: The operation of the hazardous goods containers must be complied with safety rules<br>SC2.4: The dangerous operation must be supervised by specialists<br>SC2.5: The workers and managers must have a clear understanding of the distribution of dangerous goods in the dangerous goods yard<br>SC2.6: The emergency plans of Ruihai Company should be formulated and regularly practiced<br>SC2.7: The fire fighting access must be kept unobstructed at all times<br>Business Qualification and Permission<br>SC2.8: Applying for business license for hazardous goods warehousing business and dangerous goods business must be in accordance with laws and regulations<br>SC2.9: The hazardous goods business must be carried out in accordance with laws<br>SC2.10: The renovation projects must be carried out in accordance with laws and regulations<br>SC2.11: Transforming ordinary warehouse into dangerous chemical warehouse must be obtained the governments permission<br>SC2.12: The approvals of hazardous chemicals business must be obtained legally<br>Safety Information Management<br>SC2.13: An emergency response system must be established in Ruihai Company<br>SC2.14: The major hazard sources must be identified, registered and reported to the government administrative departments.<br>**Government Safety Administration**<br>Supervision and Approval of Administrative Licenses<br>SC2.15: The approvals must be accorded with the approval procedures of governments<br>SC2.16: The activities of review and check must be according to the procedures and laws<br>SC2.17: Examining and inspection must be accorded with the approval procedures<br>Fire fighting and Emergency Rescue<br>SC2.18: The fire commander must know what the fire substances are and their characteristics and quantities before taking fire fighting measures<br>SC2.19: The fire commander must take the correct way to extinguish the fire according to the characteristics of fire substance<br>**Agency Safety Service**<br>Design and Construction Service<br>SC2.20: The design of dangerous goods yard reconstruction project must follow the overall urban planning of Tianjin and the regulatory detailed planning of Binhai New Area<br>SC2.21: The design of dangerous goods yard reconstruction project must comply with relevant safety laws<br>SC2.22: The construction of dangerous goods yard reconstruction project must comply with relevant safety laws<br>Safety Assessment Service<br>SC2.23: The report contents of safety pre-evaluation and safety acceptance evaluation must be true and reliable<br>SC2.24: The conclusion of safety pre-evaluation and safety acceptance evaluation must reflect the real situation of Ruihai Company<br>SC2.25: Safety facilities acceptance review should be comply with relevant safety laws<br>Social Safety Surveillance<br>SC2.26: Citizens and social organizations should report the illegal activities of Ruihai Company that violate the laws and regulations in time<br>SC2.27: The medium should report the illegal phenomena of enterprises in time<br>**Cyber-physical factors in Technical System**<br>SC2.28: The packaging design of nitrocellulose must be reasonable and advanced<br>SC2.29: The full range and full-time monitoring system for the hazardous chemicals must be equipped and worked well in Ruihai Company<br>SC2.30: The storage yard in Ruihai Company must be equipped with abnormal condition monitoring and alarm system<br>SC2.31: The national regulatory information system for hazardous chemicals must be built for safety administration and supervision of hazardous chemicals | **Organizational factors in Social System**<br>**Ruihai Company Safety Management**<br>Supplier Management<br>E2.1: Poor supplier management of hazardous goods<br>Process and Personal Safety Management<br>E2.2: Lack of safety training and education<br>E2.3: Failure to implement safe operation rules<br>E2.4: Lack of on-site supervision<br>E2.5: Lack of knowledge about what substances are on fire<br>E2.6: Lack of emergency evacuation training of the workers and the surrounding residents<br>E2.7: Blocked fire access<br>Business Qualification and Permission<br>E2.8: The acquisition of government' approval through illegal means<br>E2.9: Long term illegal operation of dangerous chemicals<br>E2.10: The construction of the renovation projects without approvals<br>E2.11: Changing the use of storage yard without the permission of the storage of hazardous chemicals<br>E2.12: Passing administrative examination and approval through false application materials<br>Safety Information Management<br>E2.13: Lack of access to inform the persons around the company to emergency evacuate<br>E2.14: No information on the hazardous chemicals in Ruihai Company supplied to the fire fighting officials<br>**Government Safety Administration**<br>Supervision and Approval of Administrative Licenses<br>E2.15: Approval through unconventional procedure<br>E2.16: Careless review without on-sit inspection<br>E2.17: Irresponsible on-site inspection and examining<br>Fire fighting and Emergency Rescue<br>E2.18: Lack of information on the type and quantity of fire materials on site<br>E2.19: Fire fighting by water guns in the absence of information on the fire substances<br>**Agency Safety Service**<br>Design and Construction Service<br>E2.20: The renovation project design goes against the overall planning and the detailed planning<br>E2.21: The design of a dangerous goods yard reconstruction project that violated safety laws<br>E2.22: Illegal construction of hazardous chemicals storage yard for Ruihai Company<br>Safety Assessment Service<br>E2.23: False description of key issues of Ruihai Company in the assessment report<br>E2.24: The conclusions of the assessment report violate the legal requirements<br>E2.25: Safety facilities acceptance review violates the review rules<br>Social Safety Surveillance<br>E2.26: No one or organization reports the illegal activities of Ruihai Company before the accident occurred<br>E2.27: No media report on the illegal phenomenon of Ruihai Company before the accident occurred<br>**Cyber-physical factors in Technical System**<br>E2.28: Improper packaging design of nitrocellulose<br>E2.29: The failure of hazardous goods monitoring system<br>E2.30: Lack of the alarm system<br>E2.31: Lack of real-time regulatory information system for hazardous chemicals in government departments |

**Fig. S2. Violation of safety constraints and adverse events in the meso-level analysis**

| Violation of safety constraints in the hazard-target system at the macro-level analysis | The corresponding adverse events in the hazard-target system at the macro-level analysis |
|---|---|
| **Sociological factors in Social System**<br>**Culture**<br>Government Culture<br>SC3.1: A good organizational culture must be built in government, which both written review and on-site inspection can be implemented before the approvals<br>Corporate Culture<br>SC3.2: Safety must be viewed as a core value in company<br>SC3.3: The leader at all organizational levels must have strong leaderships of safety<br>SC3.4: Everyone must clearly understand their roles with regard to process safety<br>**Law, Regulation, Rule and Standard**<br>National Law, Regulation and Standard<br>SC3.5: A unified special law on the safety management of hazardous chemicals must be established in China<br>SC3.6: The regulations on the safety management of hazardous chemicals must be specific at all stages<br>Government Regulation<br>SC3.7: Administrative staff in government should strictly enforce the law in accordance with legal procedures<br>SC3.8: Government officials should not be allowed to set administrative permits outside the law<br>SC3.9: The officials should be not allowed to interfere with the normal approval process<br>Corporate Regulation and Rule<br>SC3.10: The administrative approval of hazardous goods business must be obtained in according with the law<br>SC3.11: Everyone in company from the top to the frontline should be responsible and accountable for their roles with regard to safety<br>**Policy**<br>Government Policy<br>SC3.12: The officials and their relatives in government should not engage in commercial activities related to supervision<br>SC3.13: Information sharing channels of hazardous chemicals should be established between enterprises and government departments<br>SC3.14: The functions of government administration and enterprise operation must be separated<br>SC3.15: Central government should establish a unified system or department to ensure coordination among various relevant authorities and bodies<br>SC3.16: Local government must build the joint examination and approval system of government departments for construction projects in high-risk industries<br>SC3.17: Social surveillance plays an important role in accident prevention of enterprise<br>Corporate Policy<br>SC3.18: The information on major hazard installations in company should be made public<br>SC3.19: The workers should have the real freedom to voice their safety concerns<br>SC3.20: The enterprise should not require the evaluation report submitted by an intermediary institution to pass the reviews of the government<br>**Economics**<br>SC3.21: The cost of violating the regulations and rules must be expensive<br>SC3.22: Intermediary service agencies shall not violate the law for the sake of economic interests<br>**Technological factors in Technical System**<br>SC3.23: The storage and packaging of hazard goods must not be applied in backward technology prohibited by the law<br>SC3.24: New technologies should be applied in the national regulatory information system for hazardous chemicals and the enterprise control system to meet the statutory requirements of safety management and emergency response | **Sociological factors in Social System**<br>**Culture**<br>Government Culture<br>E3.1: Bad government culture<br>Corporate Culture<br>E3.2: Benefit rather than safety is viewed as the core value of company<br>E3.3: Managers and workers were accustomed to violate the safety rules to pursue efficiency and benefits<br>E3.4: The workers did not clearly understand their roles with regard to process safety<br>**Law, Regulation and Standard**<br>National Law, Regulation and Standard<br>E3.5: Lack of a unified special law on the safety management of hazardous chemicals in China<br>E3.6: The regulations on the safety management of hazardous chemicals in logistics enterprises is deficiency<br>Government Regulation<br>E3.7: Some government employees and officers were corrupt for the illegal approvals<br>E3.8: Some officials in government carry out administrative permission beyond the law<br>E3.9: The normal procedure of the approvals was disturbed by some officials in government<br>Corporate Regulation and Rule<br>E3.10: The actual controllers of Ruihai Company obtained the approval of hazardous chemical business through bribery and false information<br>E3.11: Violation of safety rules occurred frequently in Ruihai Company<br>**Policy**<br>Government Policy<br>E3.12: One of the actual controllers of Ruihai Company is the son of the former director of Tianjin Port Public Security Bureau<br>E3.13: The fire department can not obtain important information on the distribution and types of hazardous chemicals from Ruihai Company<br>E3.14: Tianjin Port (Group) Co., Ltd. has both the function of government supervision and the nature of enterprise management<br>E3.15: Government departments lack information sharing and joint law enforcement actions<br>E3.16: Repeated supervisions and/or regulatory loopholes exist in the safety supervision of government departments<br>E3.17: Absence of social surveillance in accident prevention<br>Corporate Policy<br>E3.18: Ruihai Company concealed the existence of major hazard installations for hazardous chemicals<br>E3.19: No one responded to the workers' safety concerns in Ruihai Company<br>E3.20: Ruihai Company required that the evaluation report of intermediary service agents must pass the approval of governments<br>**Economics**<br>E3.21: The cost of violating the regulations in China is very low<br>E3.22: Intermediary service agencies violated the legal requirements for the sake of benefits when providing technical services for Ruihai Company<br>**Technological factors in Technical System**<br>E3.23: The nitrocellulose goods were packaged into plastic bags, which were only tied up with strings and placed inside paper tubes<br>E3.24: Due to the outdated technologies, government departments can't conduct real-time supervision on hazardous chemicals and the emergency response of enterprises are often delayed |

**Fig. S3. Violation of safety constraints and adverse events in the macro-level analysis**

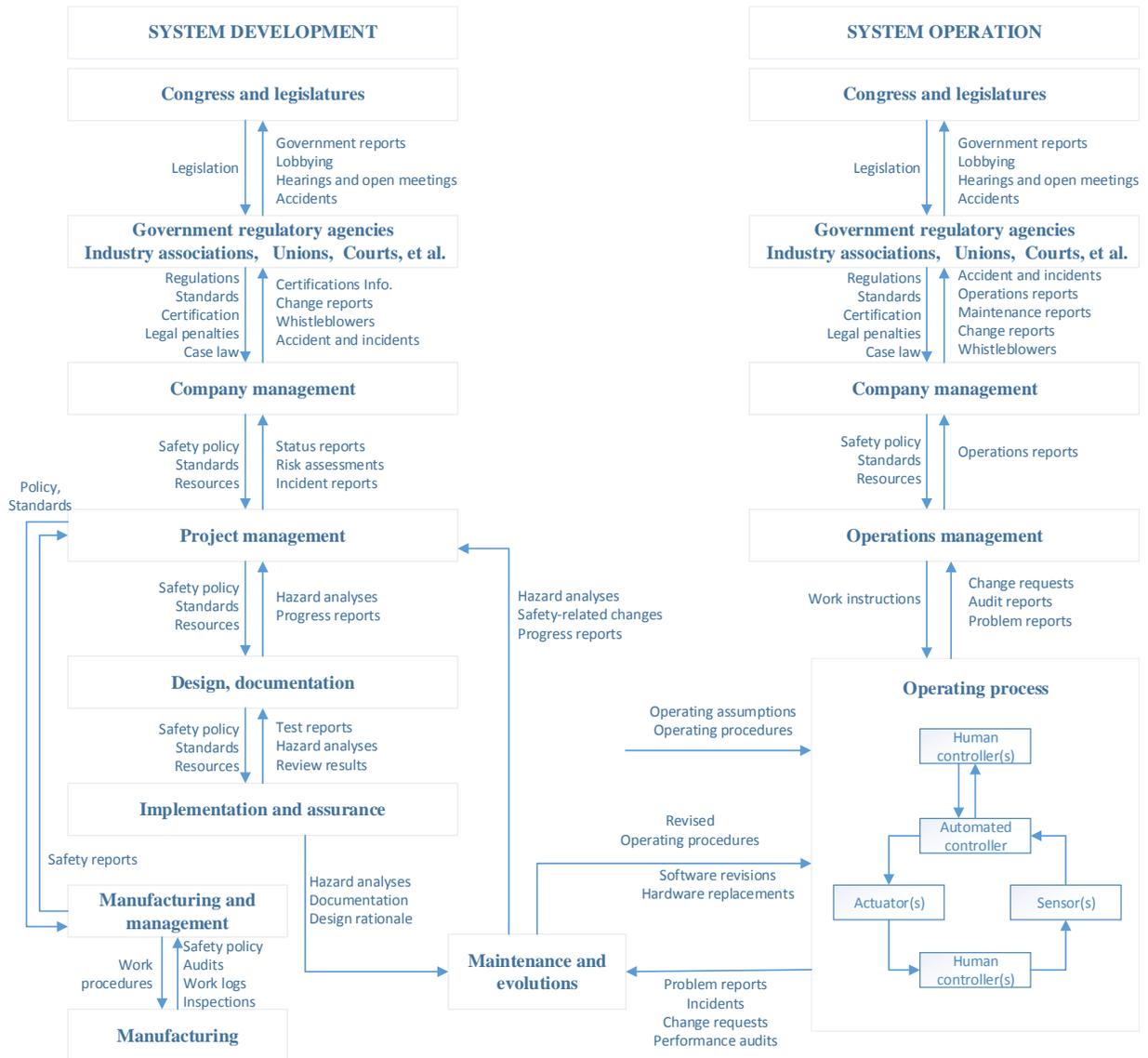

**Fig. S4. General form of the safety control structure of a sociotechnical system (*9*)**

**Table S1. Violations of safety requirements of Ruihai Company safety management**

| Safety Management of Ruihai Company | Safety-related requirements and constraints |
|---|---|
| Supplier Management Process and Personal Safety Management | The supplier management must follow the safety requirements. The workers must be trained and educated well with safety skills and safety knowledge. Handling of the hazardous goods containers must comply with safety rules. Dangerous operations must be supervised by specialists. The workers and managers must have a clear understanding of the distribution of dangerous goods in the dangerous goods yard. The emergency plans of Ruihai Company should be formulated and regularly practised. The fire passage must be kept unobstructed at all times. |
| Business Qualification and Permission | Application for a business licence for hazardous goods warehousing business and dangerous goods business must be in accordance with laws and regulations. The hazardous goods business must be carried out in accordance with laws. Renovation projects must be conducted in accordance with laws and regulations. An ordinary warehouse can be transformed into a dangerous chemical warehouse only with the government's permission. Approval of the hazardous chemicals business must be obtained legally. |
| Safety Information Management | An emergency response system must be established by Ruihai Company. The major hazard sources must be identified, registered and reported to the government administrative departments. |

**Table S2. Adverse events at the macro-level and their corresponding events at the meso- and micro-level**

| Adverse events at the macro-level analysis | The corresponding adverse events at the meso-level analysis | The corresponding adverse events at the micro-level analysis |
|---|---|---|
| $E_{3.1}$ | $E_{2.15}\ E_{2.16}\ E_{2.17}$ | $E_{1.5}$ |
| $E_{3.2}$ | $E_{2.1}\ E_{2.8}\ E_{2.9}\ E_{2.10}\ E_{2.11}\ E_{2.12}\ E_{2.28}$ | $E_{1.6}\ E_{1.9}\ E_{1.11}\ E_{1.12}\ E_{1.13}$ |
| $E_{3.3}$ | $E_{2.2}\ E_{2.3}\ E_{2.4}\ E_{2.5}\ E_{2.6}\ E_{2.7}\ E_{2.28}$ | $E_{1.7}\ E_{1.11}\ E_{1.12}$ |
| $E_{3.4}$ | $E_{2.3}, E_{2.4}\ E_{2.5}, E_{2.6}, E_{2.7}\ E_{2.13}$ | $E_{1.7}\ E_{1.8}\ E_{1.10}\ E_{1.12}\ E_{1.14}$ |
| $E_{3.5}$ | $E_{2.8}\ E_{2.9}$ $E_{2.10}\ E_{2.11}\ E_{2.12}\ E_{2.15}\ E_{2.28}$ $E_{2.29}\ E_{2.30}\ E_{2.31}$ | $E_{1.14}$ |
| $E_{3.6}$ | $E_{2.9}\ E_{2.10}\ E_{2.11}\ E_{2.14}\ E_{2.31}$ | $E_{1.12}\ E_{1.14}$ |
| $E_{3.7}$ | $E_{2.8}\ E_{2.15}$ | |
| $E_{3.8}$ | $E_{2.8}\ E_{2.15}$ | |
| $E_{3.9}$ | $E_{2.8}\ E_{2.15}$ | |
| $E_{3.10}$ | $E_{2.8}\ E_{2.12}\ E_{2.15}\ E_{2.23}\ E_{2.24}\ E_{2.25}$ | $E_{1.14}$ |
| $E_{3.11}$ | $E_{2.1}\ E_{2.2}\ E_{2.3}\ E_{2.4}\ E_{2.5}\ E_{2.6}\ E_{2.7}$ | $E_{1.6}\ E_{1.7}\ E_{1.9}\ E_{1.11}\ E_{1.12}$ |
| $E_{3.12}$ | $E_{2.8}\ E_{2.9}\ E_{2.15}$ | |
| $E_{3.13}$ | $E_{2.14}\ E_{2.18}\ E_{2.19}$ | |
| $E_{3.14}$ | $E_{2.8}\ E_{2.9}$ $E_{2.10}\ E_{2.12}\ E_{2.15}\ E_{2.16}\ E_{2.17}$ | $E_{1.5}$ |
| $E_{3.15}$ | $E_{2.8}\ E_{2.9}\ E_{2.10}\ E_{2.11}\ E_{2.12}\ E_{2.18}$ | |
| $E_{3.16}$ | $E_{2.8}\ E_{2.9}\ E_{2.10}\ E_{2.11}$ | |
| $E_{3.17}$ | $E_{2.26}\ E_{2.27}$ | |
| $E_{3.18}$ | $E_{2.18}\ E_{2.19}$ | $E_{1.13}$ |
| $E_{3.19}$ | $E_{2.5}\ E_{2.6}$ | |
| $E_{3.20}$ | $E_{2.23}\ E_{2.24}$ | |
| $E_{3.21}$ | $E_{2.8}\ E_{2.9}$ $E_{2.10}\ E_{2.11}\ E_{2.12}\ E_{2.20}\ E_{2.21}$ $E_{2.22}\ E_{2.23}\ E_{2.24}\ E_{2.25}$ | $E_{1.5}$ |
| $E_{3.22}$ | $E_{2.20}\ E_{2.21}\ E_{2.22}\ E_{2.23}\ E_{2.24}\ E_{2.25}$ | $E_{1.5}$ |
| $E_{3.23}$ | $E_{2.1}\ E_{2.28}\ E_{2.29}\ E_{2.30}$ | |
| $E_{3.24}$ | $E_{2.13}\ E_{2.14}\ E_{2.18}\ E_{2.29}\ E_{2.30}\ E_{2.31}$ | $E_{1.13}$ |

**Table S3. List of recommendations**

| Controllers | Recommendations |
|---|---|
| The national legislative and policymaking body | Unified laws on the safe management of hazardous chemicals should be established by the National People's Congress (NPC) in China |
| | Specific requirements for circulation and use of hazardous chemicals should be added to the regulations on safe management of hazardous chemicals |
| | A national regulatory information system for hazardous chemicals must be built for safe administration and supervision of hazardous chemicals |
| Government and regulation body | **Government Culture** |
| | A good organizational culture must be built in the government, with both written review and on-site inspection implemented before approval |
| | **Government Regulation** |
| | Administrative staff in government should strictly enforce the law in accordance with legal procedures |
| | Government officials should not be allowed to grant administrative permits outside the law |
| | Officials should be not allowed to interfere with the normal approval process |
| | **Government Policy** |
| | Officials and their relatives in government should not engage in commercial activities related to supervision |
| | Information sharing channels of hazardous chemicals should be established between enterprises and government departments |
| | The functions of government administration and enterprise operation must be separated |
| | The central government should establish a unified system or department to ensure coordination among various relevant authorities and bodies |
| | Local governments must build a joint examination and approval system of government departments for construction projects in high-risk industries |
| | Social surveillance should play an important role in the accident prevention of an enterprise |
| Corporate (Enterprises engaged in storage and circulation of hazardous chemicals) | **Corporate Culture** |
| | Safety must be viewed as a core value by the company |
| | The leader at all organizational levels must have strong safety leadership |
| | Everyone must clearly understand their roles with regard to process safety |
| | **Corporate Regulation and Rules** |
| | The administrative approval of hazardous goods business must be obtained in accordance with the law |
| | Everyone in the company from the top to the frontline should be responsible and accountable for their roles with regard to safety Policy |
| | **Corporate Policy** |
| | Information about major hazard installations should be made public |
| | Workers should have the freedom to voice their safety concerns |
| | The enterprise should require evaluation reports submitted by an intermediary institution to pass government review |
| Intermediary agents | The intermediary agencies providing safety services should not engage in illegal activities such as providing false reports in order to pursue profits |
| Social organizations and media | Citizens, social organizations and the media should report to the government departments immediately when a safety violation is identified in a high-risk industrial company |
| Technical aspects | New technologies should be applied in the national regulatory information system for hazardous chemicals and the enterprise control system to meet the statutory requirements of safety management and emergency response |